\documentclass{article}

\usepackage{PRIMEarxiv}

\usepackage[utf8]{inputenc} 
\usepackage[T1]{fontenc}    
\usepackage{authblk} 
\usepackage{hyperref}       
\usepackage{url}            
\usepackage{tabularx}
\usepackage{booktabs}       
\usepackage{amsfonts}       
\usepackage{nicefrac}       
\usepackage[table]{xcolor} 
\usepackage{microtype}      
\usepackage{amsmath}        
\usepackage{lipsum}
\usepackage{fancyhdr}       
\usepackage{graphicx}       
\graphicspath{{media/}}     

\usepackage{algorithm}
\usepackage{algorithmic}

\usepackage{wrapfig}
\usepackage[numbers]{natbib}
\PassOptionsToPackage{numbers, compress}{natbib}
\usepackage{amssymb}
\usepackage{multirow}
\usepackage{amsmath}
\title{BadBlocks: Low-Cost and Stealthy Backdoor Attacks Tailored for Text-to-Image Diffusion Models
}
\author[1]{\href{jiawu.sspu@gmail.com}{Jia Wu}}
\author[2]{\href{yupan.sspu@gmail.com}{Yu Pan}}
\author[1]{\href{junjunyang.sspu@gmail.com}{Junjun Yang}}
\author[1]{\href{duyi@sspu.edu.cn}{Yi Du}}
\affil[1]{%
    Shanghai Polytechnic University, Shanghai 201209, China
}
\affil[2]{%
    ShanghaiTech University, Shanghai 201210, China
}

\begin{document}
\maketitle 

\begin{abstract}
Despite the remarkable progress of diffusion models in image generation, recent studies reveal their vulnerability to backdoor attacks via covert visual or textual triggers. Although evolving defense mechanisms can detect most existing threats through visual inspection or feature analysis, we introduce BadBlocks—a novel, lightweight, and highly covert attack that challenges these safeguards. By selectively poisoning specific blocks within the UNet architecture while keeping other components intact, BadBlocks requires only 30\% of the computational resources and 20\% of the GPU time of conventional attacks, effectively democratizing backdoor injection on consumer-grade GPUs. Empirical evaluations demonstrate that BadBlocks achieves a high attack success rate with negligible perceptual quality loss, while successfully bypassing state-of-the-art defenses, particularly attention-based detection frameworks. Layer-level ablation studies further confirm that backdoor mapping does not require full-network fine-tuning, revealing the disparate vulnerability of different neural layers. Overall, BadBlocks significantly lowers the barrier for executing backdoor attacks, presenting a critical security risk. Our code is available at: \url{https://github.com/paoche11/BadBlocks}.
\end{abstract}

\section{Introduction}
\label{sec:intro}
Today, generative models have become some of the most widely adopted models in the field of artificial intelligence, with diffusion models emerging as one of the most effective approaches. Diffusion models are extensively used to generate high-quality images and videos, and they support a wide range of conditional inputs, such as text prompts, reference images, and ControlNet maps to guide the generation process\cite{diffusion_survey1,diffusion_survey2}. 

However, recent studies have shown that diffusion models are vulnerable to backdoor attacks in which attackers can manipulate the output of models by embedding covert triggers in the input\cite{trojdiff,baddiffusion}. Backdoor attacks pose significant challenges to the safe development of diffusion models. Once the backdoor is activated by attackers, it can lead to unpredictable and potentially harmful outcomes, such as the generation of malicious content or the manipulation of decisions in downstream classifiers\cite{survey,survey2}. In addition, due to the stealthy nature of backdoors and the diversity of potential triggers, backdoor attacks are widely regarded as one of the most serious threats to diffusion models. In threat scenarios, attackers upload backdoor-injected models to a public platform (e.g., Hugging Face or Github) while falsely claiming it to be benign. When unsuspecting users download and deploy these models, their output typically appears normal and harmless. However, once the attacker activates the backdoor, the target model begins to generate predefined malicious content\cite{BadT2I}.

The powerful generative capabilities of diffusion models, coupled with their vulnerability to backdoor attacks, have raised widespread concerns. With the increasing release of pre-trained models, users frequently download the necessary models from open platforms according to their specific requirements\cite{pretrained:1,pretrained:2}. However, due to the opacity of black-box models, users often focus solely on whether the pre-trained models can accomplish the intended tasks, while overlooking potential backdoor threats\cite{backdoor:1,backdoor:2,backdoor:3}. For example, in classification tasks, attackers can easily manipulate the output by activating backdoors\cite{backdoor:2}. In face recognition tasks, when facial images contain trigger patterns, the model may misclassify them as any identity predetermined by the attacker\cite{faceback}. In generative tasks, backdoor attacks often cause the model to produce malicious content, including violent, gory, or otherwise inappropriate imagery, which may further compromise downstream applications\cite{llmback}. In diffusion models, the objective of backdoor attacks is often to generate a specific image or a particular category of images.

Fortunately, successfully executing covert backdoor attacks remains a highly challenging task. On the one hand, various defense frameworks have been developed to counter backdoor attacks in diffusion models. These frameworks typically achieve backdoor detection and trigger inversion by designing specialized neural network architectures and tailored loss functions\cite{eliagh,terd}. On the other hand, attackers often incur significant costs for computing resources and must invest considerable time in fine-tuning the model to inject covert and effective backdoors, which limits the scalability of backdoor attacks.

However, in this paper, we find that the influence of backdoor attacks on the parameters of the diffusion model can be modeled as fine-grained sample blocks. Based on this discovery, we reveal a novel threat that enables attackers to inject backdoors into diffusion models with minimal cost and significantly reduced training time.  Notably, BadBlocks exhibits strong stealthiness and effectively mitigates the assimilation phenomenon in attention layers.  Our main contributions are summarized as follows:
\begin{itemize}
\item \textbf{We propose BadBlocks}, a lightweight, stealthy, and fine-grained backdoor attack framework that injects backdoors into diffusion models by optimizing only the most vulnerable architectural blocks. Empirical evaluations demonstrate that BadBlocks reduces video memory usage by up to 70\% and GPU training time by 80\% compared to baseline approaches, democratizing the attack execution on standard consumer-grade graphics cards.

\item \textbf{We identify and model parameter sensitivity profiles} across different neural network layers in diffusion models for the first time, revealing that varying components exhibit disparate vulnerability degrees to backdoor injections. Furthermore, we provide an in-depth analysis of the underlying causes behind the "assimilation phenomenon" in Text-to-Image backdoors, exposing a critical blind spot in existing defense mechanisms that rely on attention-feature detection.

\item \textbf{We conduct a granular, layer-level ablation study} to dissect the structural roles of specific network components during backdoor deployment. While prior literature heavily focuses on coarse, model-level vulnerabilities, our work is the first to analyze backdoor risks at the layer level, providing a rigorous interpretability foundation for future research into diffusion safety and defense mechanisms.
\end{itemize}
\section{Related Works}
\label{sec:rw}

This section introduces the fundamentals of diffusion models, reviews representative backdoor attack and defense methods, and discusses their limitations.

\subsection{Fundamental Principles of Diffusion Models}
\label{sec:dms}

Diffusion models are powerful generative models capable of learning data distributions from dataset \(x\). \textit{Denoising Diffusion Probabilistic Models} (DDPMs) first introduced diffusion-based image generation\cite{ddpm}. DDPM consists of a forward diffusion process and a reverse denoising process. The forward process can be represented as \(x_0 \rightarrow x_t\), where
\[
x_t = \sqrt{\alpha_t}x_0 + \sqrt{1-\bar{\alpha}_t}\epsilon.
\]

The reverse process iteratively removes noise:
\[
\mathbf{x}_{t-1} =
\frac{1}{\sqrt{\alpha_t}}
\left(
\mathbf{x}_t -
\frac{1-\alpha_t}{\sqrt{1-\bar{\alpha}_t}}
\mathbf{\epsilon}_\theta(\mathbf{x}_t,t)
\right)
+
\sigma_t\epsilon,
\quad
\epsilon \sim \mathcal{N}(\mathbf{0}, \mathbf{I}),
\]
where \(\epsilon_\theta\) denotes the predicted noise.

\textit{Denoising Diffusion Implicit Models} (DDIMs) accelerate sampling by removing Markovian dependency in the reverse process\cite{ddim}. \textit{Latent Diffusion Models} (LDMs) further reduce computational costs by compressing images into latent space using a Variational Autoencoder (VAE)\cite{stablediffusion,vae}. \textit{Stochastic Differential Equation} (SDE)-based methods unify diffusion processes with stochastic differential equations and optimize the score function \(\nabla_x \log P_t(x)\)\cite{sde}. In addition, ODE-based schedulers further reduce sampling steps to 15--30\cite{dpmv1,dpmv3}.

\subsection{Backdoor Attacks in Diffusion Models}
\label{sec:ba}

Backdoor attacks are considered a major security threat to diffusion models\cite{riskindiff}. Attackers establish covert neuron mappings through fine-tuning, enabling trigger inputs to produce predefined outputs\cite{backdoor:2}. TrojDiff was the first method to inject trigger-specific perturbations into the noise space of DDPM and DDIM models\cite{trojdiff}. BadDiffusion directly injects patches into the noise space without relying on trigger embedding functions\cite{baddiffusion}. RickRolling further extends the attack space to textual prompts, enabling attacks on LDMs such as Stable Diffusion\cite{rickroll}. VillanDiffusion unifies text- and noise-based attack spaces within a single framework\cite{villan}. Additional attack spaces have also been explored. For example, Gungnir employs image style as a backdoor trigger in image-to-image tasks\cite{gungnir}. These studies demonstrate the continuous expansion of backdoor attack spaces in diffusion models.

\subsection{Defense Strategies in Diffusion Models}
\label{sec:ds}

As backdoor threats become increasingly severe, various defense strategies have been proposed. Elijah is the first framework specifically designed to defend against diffusion-model backdoor attacks and can mitigate BadDiffusion, TrojDiff, and VillanDiffusion attacks\cite{eliagh}. TERD further formulates noise-space attacks as
\[
x_t = a(x_0,t)x_0 + b(t)\epsilon + c(t)r,
\]
where \(a\) and \(b\) denote image-noise mixing functions and \(c\) represents trigger injection\cite{terd}, which detects backdoors by measuring the KL divergence between inverted and benign noise distributions.

For prompt-based attacks, T2IShield is the first defense framework targeting textual triggers\cite{T2IShield}. By analyzing attention maps and identifying the ``Assimilation Phenomenon,'' T2IShield effectively mitigates attacks such as RickRolling. Although defense methods have evolved from handcrafted architectures to distribution-based black-box analysis, many attack spaces remain underexplored, including ControlNet and adapter-based tasks\cite{IP-Adapter}.

\subsection{Limitations}
\label{sec:lim}

Backdoor injection in diffusion models faces two major challenges. First, it requires substantial computational resources, particularly GPU memory, since attacking diffusion models can be nearly as expensive as full-scale training. Second, target-specific poisoning often causes the ``assimilation phenomenon,'' increasing the risk of detection.

Existing defense methods also lack generalization across attack spaces. Elijah and TERD mainly focus on noise-space attacks, while T2IShield is limited to prompt-based attacks.

\section{Method}
\label{sec:meth}
In this section, we introduce the BadBlocks threat model, outlining its attack scenarios and background. We analyze the capabilities of both attackers and defenders, and briefly describe the attack method with a theoretical explanation of its feasibility.
\subsection{Threat Model}
\label{sec:tm}
The threat model for backdoor attacks on diffusion models is typically defined as a black-box setting from the user's perspective. Depending on the attack method, adversaries can poison the training data or alter the training strategy, ultimately providing backdoored models to the user. The threat model in BadBlocks is largely consistent with previous work. However, because the weight distribution and features of the remaining blocks remain unchanged, BadBlocks is more likely to evade detection by defense frameworks, especially which methods based on the global features of the UNet structure. 
\begin{figure}[htbp]
    \centering
    \includegraphics[width=\linewidth]{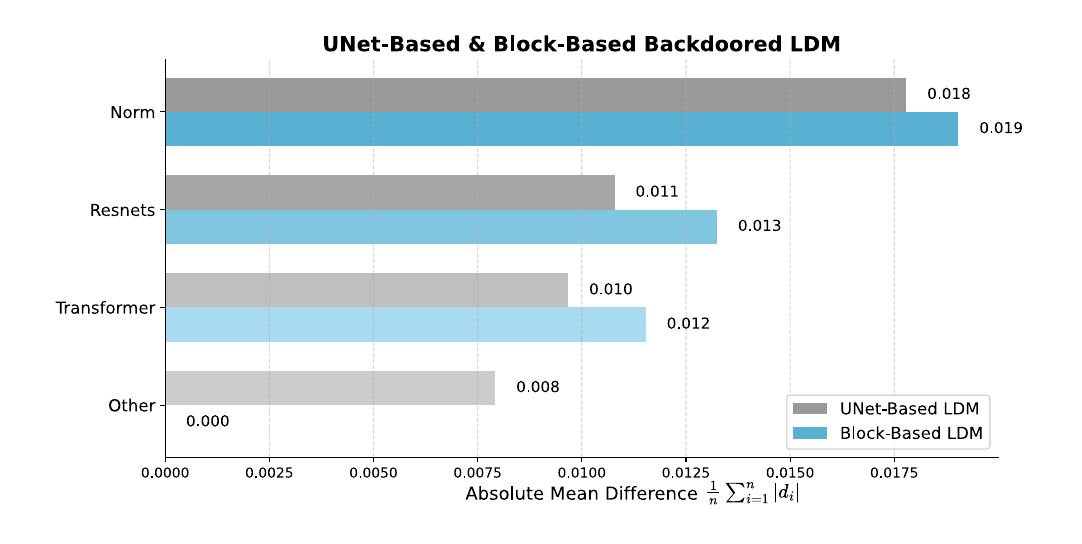}
    \caption{In infected UNet models, the normalization layers exhibit the most significant weight changes, followed by the ResNet layers and Transformer blocks. We hypothesize that this hierarchy of changes is critical for establishing effective backdoor mappings.}
    \label{fig:fig3}
\end{figure}
\vspace{-20pt}
\subsection{Attacker and Defender Knowledge}
\label{sec:know}
Building on existing work\cite{BadNL,DBA}, we define the knowledge of the attacker in the BadBlocks setting as follows: (1) The attacker has the authority to manipulate the fine-tuning process of the model, including the dataset, the loss function and the training strategy; (2) The attacker can access the model and provide input data that comply with its specifications. In contrast, a model defender not only has full access to all model parameters and can input arbitrary data to obtain detection samples, but can also perform additional operations such as distillation, pruning, and other modifications about model weights \cite{distillerase,diff-clean}.
\subsection{Mitigation Assimilation}
\label{sec:ea}
The key intuition behind the design of BadBlocks is that the assimilation phenomenon is closely tied to the cross-attention layers across various blocks in the diffusion model. Therefore, to effectively eliminate assimilation, it is crucial to preserve benign weights in as many attention layers as possible. This raises two key questions: (1) Can the impact of backdoor-infected weights be minimized? (2) Are all UNet blocks and neural network layers essential for executing a backdoor attack?

Inspired by previous studies\cite{hm,le}, we recognize that different neural network blocks within the UNet architecture of a diffusion model serve distinct roles during the image generation process. Therefore, we attempt to design an algorithm that decouples the neural components associated with backdoor mapping from the UNet and applies fine-grained training to them. Specifically, in the attention-based detection method, given a token sequence of length $L$, each block $b$ generates a cross-attention map $M_b^{(t)}$ at any time step $t \in T$:
\begin{equation}
M_b^{(t)} = \text{CrossAttn}(Q^{(t)}, K^{(t)}, V^{(t)}) \in \mathbb{R}^{L \times D},
\end{equation}
\(T\) is a hyperparameter that determines the number of iterations of sampling performed by the model. We denote the network depth of UNet as \(D\), representing all the sampling blocks in the network. By averaging the attention maps throughout the sampling process, we obtain the attention map sample \(\hat{M}\) used for backdoor detection:
\begin{equation}
\bar{M}_b = \frac{1}{|T|} \sum_{t = 1}^{T} M_b^{(t)},
\end{equation}
\begin{equation}
\hat{M} = \frac{1}{|D|} \sum_{b = 1}^{D} \bar{M_b}.
\end{equation}

In the defense evaluation of BadBlocks, we adopt the F-Norm Threshold Truncation method proposed in T2IShield, a lightweight detection approach grounded in the assimilation phenomenon. Specifically, by analyzing the F-Norm distribution, we derive the detection sample F:
\begin{equation}
    F = \| \hat{M} \|_F = \frac{1}{L}\sum_{t=1}^{L}\sqrt{ \sum_{i=1}^{D} \sum_{j=1}^{D} (\bar{M_{b}}-\hat{M})^2},
\end{equation}
Finally, a threshold value \(\hat{F} \in \mathbb{R}^{1}\) is applied to determine whether the sample contains a backdoor trigger:
\begin{equation}
\textit{Sample is} =
\begin{cases}
\text{benign}, & \text{if } F \geq \hat{F}, \\
\text{backdoor}, & \text{if } F < \hat{F}.
\end{cases}
\end{equation}

However, in BadBlocks, the attention maps \(M_b\) generated by most blocks appear to be entirely normal, as their parameters remain unchanged. Only the injected sampled blocks that are targeted by the attack exhibit abnormal attention patterns, and these anomalies can be easily masked when averaged with the attention maps from benign blocks. This allows the remaining benign cross-attention layers to suppress the assimilation effect induced by the backdoor. Ultimately, through controlled experiments, we successfully isolated the network blocks essential for executing backdoor attacks in diffusion models, specifically, network blocks in the upsampling stage and its critical subcomponents: the \textbf{the resnets}\cite{resnet}, \textbf{the transformer blocks}\cite{trans} and \textbf{the normalization layers}\cite{groupnorm} among them.
\renewcommand{\arraystretch}{1.8}
\begin{table*}[t]
\centering
\caption{We evaluated the minimum computational overhead of BadBlocks for 20,000 samples on an NVIDIA A40 GPU. Compared to existing methods, BadBlocks introduces significantly less impact on the target model. Moreover, its low dependency on attacker-side hardware substantially lowers the barrier to executing backdoor attacks.}
\label{tab:tab1}
\resizebox{\textwidth}{!}{ 
\begin{tabular}{|c|cccc|}
\hline
\textbf{Method}          & \textbf{Infected Components} & \textbf{Affected Parameters} & \textbf{Necessary GPU Memory} & \textbf{GPU Time (1 epoch)} \\ \hline
\textit{BadT2I}          & \textbf{UNet}                & \textbf{859M/1066M}          & \textbf{18529MB}              & \textbf{7.5 hour}           \\ \hline
\textit{VillanDiffusion} & \textbf{UNet \& Text Encoder} & \textbf{1066M/1066M}         & \textbf{21425MB}              & \textbf{9.5 hour}           \\ \hline
\textit{RickRolling}     & \textbf{Text Encoder}        & \textbf{123M/1066M}          & \textbf{6971MB}               & \textbf{3.2 hour}           \\ \hline
\textit{BadBlocks(Ours)} & \textbf{UpSample Blocks}     & \textbf{18M / 1066M}         & \textbf{5589MB}               & \textbf{1.4 hour}           \\ \hline
\end{tabular}
}
\end{table*}
\begin{figure}[htbp]
    \centering
    \includegraphics[width=1\linewidth]{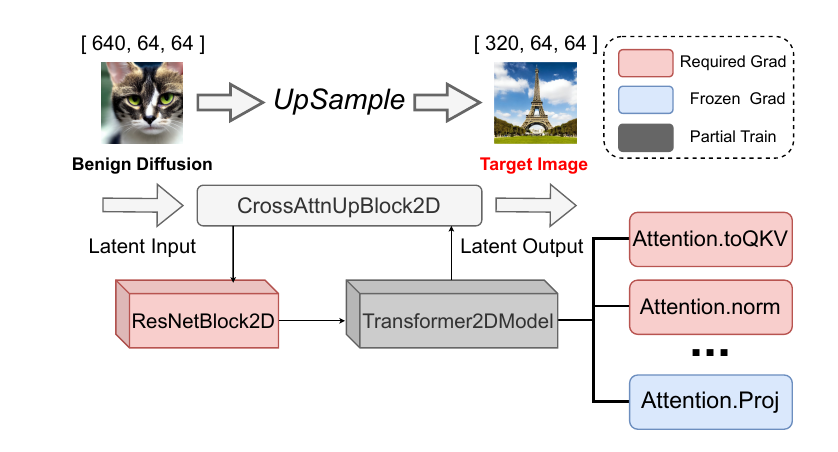}
    \caption{Our findings indicate that the final upsampling block plays a critical role in enabling backdoor mapping, while also demonstrating that not all model parameters are essential for executing backdoor attacks.}
    \label{fig:fig4}
\end{figure}
\vspace{-20pt}
\subsection{Approaches of BadBlocks}
\label{sec:appro}
The training of diffusion models typically involves a model \(M\) parameterized by weights \(\theta\), which learns to approximate the data distribution from the original dataset \(x\). The trained model is capable of generating image \(x'\) from noisy inputs \(\epsilon\) conditioned on additional information \(a\), which can be formally expressed as:
\begin{equation}
    x' = M_{\theta}(\epsilon,a,t),\epsilon\sim{N(0, I)}, 
\end{equation}
where \(t\in{T}\) is the variance of timestep.

We adopt the definition of the input space \(S_{input}\) from Gungnir\cite{gungnir}, where the additional conditions space \(a\subseteq{A_{cond}}\) is introduced to constrain the stochasticity of the diffusion process. The complete composition of the additional space \(A_{cond}\) is:
\begin{equation}
    A_{cond}=\{prompts,images,controlnet,...\}.
\end{equation}
Therefore, the complete input space \(S_{input}\) for the diffusion process can be defined as follows:
\begin{equation}
     S_{input}=\{(\epsilon,a)|\epsilon\sim{N(0,I)},a\subseteq{A_{cond}}\}.
\end{equation}

In BadBlocks, we define a hyperparameter \(i\geq0\) to represent the number of blocks involved in backdoor training, where \(i\) is strictly less than the total number of upsample blocks. Importantly, during training, we froze nearly all parameters of the model \(M_{\theta}\), performing gradient updates only on the critical weight \(\theta^*_i\), including the ResNet layers, Transformer blocks, and normalization layers. Figure~\ref{fig:fig4} illustrates the specific parameters within the sampled blocks targeted by BadBlocks that require training.

\begin{algorithm}[!ht]
    \renewcommand{\algorithmicrequire}{\textbf{Input:}}
    \renewcommand{\algorithmicensure}{\textbf{Output:}}
    \caption{Overall BadBlocks Training Procedure}
    \begin{algorithmic}[1]
        \REQUIRE  Target model $M$, Clean dataset $\mathbf{D_{c}}$, Secret trigger $\mathbf{s}$, Backdoor target $\mathbf{i_t}$, Number of BadBlocks $i$ \,Model parameters $\theta$, Poisoning rate $r$, Learning rate $\eta$;
        \ENSURE Backdoored model $M^{'}_{\theta^{*}}$;
        \STATE $\mathbf{D_p} \leftarrow \text{Poison}(\mathbf{D_c}, \mathbf{s}, \mathbf{i_t}, r)$; \# Generate poison dataset
        \STATE $\mathbf{D} = \{\mathbf{D_{c}}, \mathbf{D_{p}}\}$; \# Merge into training dataset
        \STATE $S = \{(\epsilon,a_b)\}$; \# Define input space
        \STATE $\text{Required\_Grad($\theta, False$) },\text{Required\_Grad(} \theta^*_i, True)$;
        \WHILE{remaining epochs}
            \STATE $x_0 \sim$ Uniform $\mathbf{D_{p}}$; 
            \IF{$backdoor \ training$}
                \STATE $t \sim$ Uniform$({1,...,T})$;
                \STATE $\mathrm{d} \mathbf{x}_t = f(t, \mathbf{x}_t) \, \mathrm{d}t + g(t) \, \mathrm{d}\mathbf{w}_t$; \# Add random noise
                \STATE $\mathcal{L} = \mathbb{E}_{x_0,s,t}\left[\left\|\epsilon-\epsilon_{\theta^{*}_{i}}(x_t,a_b,t)\right\|^{2}_2\right]$;
            \ENDIF
            \STATE \text{Update } $\theta^*_i$ \text{ using optimizer step on loss } $\mathcal{L}$
        \ENDWHILE
        \STATE \textbf{return} $M_{\theta^{*}}^{'}$; \quad \# Return the backdoored model
    \end{algorithmic}
    \label{algorithmA}
\end{algorithm}

Owing to the universality of BadBlocks, its additional condition space in backdoor attacks, denoted as \(a_b\), closely approximates the entire benign input space \(A\). This indicates that BadBlocks can accommodate nearly any existing trigger dimension. Based on this input space, we define the final training loss function for the BadBlocks backdoor as follows:
\begin{equation}
     L_{badblocks}=E_{x_0,s,t}[||\epsilon-\epsilon_{\theta^{*}_i}(x_t,a_b,t)||^2_2].
\end{equation}

\begin{figure*}[t]
    \centering
    \includegraphics[width=1.0\linewidth]{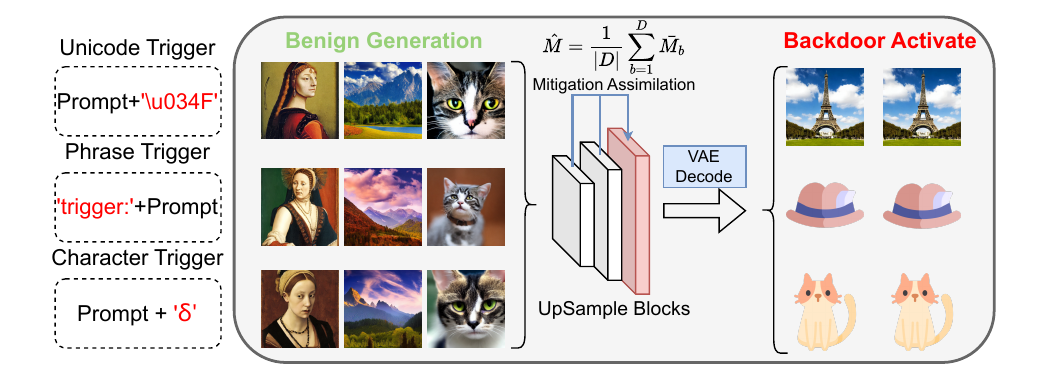}
    \caption{One key advantage of BadBlocks is that it does not rely on the modification of the loss function, enabling broad compatibility with existing backdoor methods. It consistently produces high-quality images across various triggers with minimal degradation.}
    \label{fig:fig5}
\end{figure*}

Algorithm~\ref{algorithmA} presents the fine-tuning process of BadBlocks, where the training procedure is carried out by freezing the majority of weights. Furthermore, to quantify the weight changes introduced by backdoor attacks, we compute the absolute mean difference \(\mathbf{d}\) between the weights of the benign and backdoor model. Every $d_i$ between each pair of benign and backdoor weights can be expressed as:
\begin{equation}
d_i = \frac{1}{n_i} \sum_{j=1}^{n_i} \left| \theta_{i,j}^{(benign)} - \theta_{i,j}^{(backdoor)} \right|,
\end{equation}
\begin{equation}
    \mathbf{d} = [d_1, d_2, \dots, d_k],
\end{equation}
where \(k\) is the total number of parameter tensors in the block. In the experimental section, we leverage this differential analysis to identify the parameters that play a pivotal role in establishing backdoor mappings within neural networks, thereby providing critical evidence to support our findings. Figure~\ref{fig:fig3} visualizes the differences in weight changes across components affected by BadBlocks.

\section{Experiments}
\label{sec:exper}
In this section, we present the evaluation results of BadBlocks in the text-to-image task, along with its corresponding attack settings. Figure~\ref{fig:fig5} shows the attack process of BadBlocks. Furthermore, we highlight the advantages of our method in terms of training efficiency and resource consumption while also acknowledging its limitations.
\subsection{Attack Configuration}
\label{sec:config}
In the experimental section, we adopted standard settings for the backdoor attack. For the text-to-image task, we used three different characters as the backdoor trigger, including invisible Unicode, specific phrase, and lowercase greek letter.  As attack targets, we selected three mainstream diffusion models: Stable Diffusion v1.5, Stable Diffusion v2.1 and Realistic Vision v4.0, which were trained on the MS-COCO-Caption dataset\cite{coco}. To assess the quality of the generated images, we use the FID score as the primary metric\cite{FID}. Furthermore, we employ LPIPS and SSIM to evaluate the attack success rate (ASR)\cite{LPIPS,ssim}. All training is conducted for 5 epochs, with a learning rate of \(1e-4\) and a batch size of 8. All training and evaluation processes were conducted on an NVIDIA A40 GPU, but this does not mean that professional computing graphics cards are necessary for BadBlocks. Our experimental results demonstrate that consumer-grade GPUs, such as the NVIDIA RTX 3060 (12GB) and RTX 4070Ti SUPER (16GB), can also perform efficient and successful backdoor attacks within a short time frame. Table~\ref{tab:tab1} presents a comparison between BadBlocks and existing approaches in terms of parameter modification scope and GPU resource consumption.

\subsection{Attack Results}
\label{sec:ar}
Table~\ref{tab:tab2} presents the attack results of BadBlocks on three different baseline models. The experimental results prove that establishing a backdoor mapping requires only a few key weights. In Stable Diffusion v1.5, the FID loss was merely 4.3\%, whereas the UNet-based method exhibited a loss of 73.4\% under the same poisoning rate. In the other two baselines, BadBlocks also demonstrated similar effects. The experimental results demonstrate the feasibility of implementing a fine-grained backdoor training scheme by decoupling the model architecture, indicating that backdoor attacks often require infection of only the key components. 
a
\begin{table}[t]
\centering
\renewcommand{\arraystretch}{1.4} 
\caption{We evaluated the performance of BadBlocks across three different baseline models and found that it not only executed backdoor attacks effectively, but also exhibited strong stealthiness.}
\label{tab:tab2}
\begin{tabularx}{\linewidth}{c|c|>{\centering\arraybackslash}X|>{\centering\arraybackslash}X}
\hline
\textbf{Models} & \textbf{ASR $\uparrow$} & \textbf{FID Score $\downarrow$} & \textbf{Target Similarity} \\ \hline
\multirow{2}{*}{SD v1.5} & \multirow{2}{*}{99.95\%} & Baseline: 17.98 & LPIPS: 0.16$\downarrow$ \\
& & \cellcolor[HTML]{C0C0C0}\textbf{BadBlocks: 18.78} & SSIM: 0.69$\uparrow$ \\ \hline
\multirow{2}{*}{SD v2.1} & \multirow{2}{*}{99.82\%} & Baseline: 24.74 & LPIPS: 0.13$\downarrow$ \\
& & \cellcolor[HTML]{C0C0C0}\textbf{BadBlocks: 22.23} & SSIM: 0.71$\uparrow$ \\ \hline
\multirow{2}{*}{RV v4.0} & \multirow{2}{*}{99.97\%} & Baseline: 19.97 & LPIPS: 0.03$\downarrow$ \\
& & \cellcolor[HTML]{C0C0C0}\textbf{BadBlocks: 19.37} & SSIM: 0.81$\uparrow$ \\ \hline
\end{tabularx}
\end{table}

\subsection{Advantages and Limitations}
\label{sec:andl}
Experimental results demonstrate that backdoor attacks based on diffusion models only require minimal parameter training. Effective attacks can be achieved by updating a small subset of model parameters. The primary advantage of our method lies in its ability to significantly reduce parameter modifications, thereby lowering computational costs and reducing training time. Furthermore, BadBlocks is highly compatible with other attack strategies, as its effectiveness does not rely on loss function manipulation. Additional experiments combining BadBlocks with mainstream attack methods are provided in the supplementary material, demonstrating that BadBlocks can significantly reduce resource consumption without compromising attack performance while further enhancing concealment. Despite the demonstrated efficiency of BadBlocks, certain limitations remain. Experimental results indicate that the poisoning rate (PR) significantly affects the quality of the generated target images. When the PR falls below 10\%, the target model is still capable of producing backdoor images. However, due to the limited reconstruction capability of a single sampled block, the model struggles to generate high-resolution targets. This constraint imposes certain limitations on the types of datasets that attackers can effectively exploit. Detailed attack results under different poisoning rates across three baseline models are provided in the supplementary material.

\section{Ablation Study}
\label{sec:abs}
In this section, we investigate the impact of different configurations on the attack performance, including ablations of key components and the different number of infected blocks.
\subsection{Impact of Crucial Components}
\label{sec:cc}
\begin{figure}[htbp]
    \centering
    \includegraphics[width=1\linewidth]{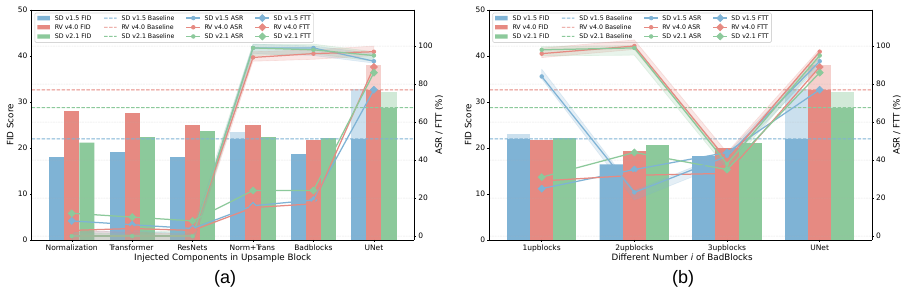}
    \caption{Figure~\ref{fig:fig7}(a) shows that a single component is insufficient for backdoor mapping, and effective attacks require at least the ResNet and Transformer blocks. Figure~\ref{fig:fig7}(b) shows that BadBlocks performance correlates with the number of infected blocks. It is evident that the attack performance is not positively correlated with the number of affected parameters, suggesting that certain components within the model are particularly sensitive to backdoor injection.}
    \label{fig:fig7}
    \vspace{-5px}
\end{figure}

To meticulously dissect the mechanistic contributions of distinct structural components during backdoor deployment and systematically minimize the adversarial attack surface, we performed a series of granular ablation studies focusing on the ResNet layers, Transformer blocks, and normalization layers embedded within the core sampling blocks. As quantified by our empirical results, isolated single-component fine-tuning is inherently insufficient to anchor the backdoor trigger mapping within the network. Conversely, jointly optimizing the normalization layers and Transformer blocks successfully facilitates backdoor injection; however, this restricted optimization comes at the expense of a non-negligible degradation in original image synthesis quality. Crucially, as illustrated in Figure~\ref{fig:fig7}(a), although the ResNet layer does not actively participate in the explicit expression or retrieval of the backdoor payload, it plays an indispensable role as a structural stabilizer. It serves to preserve the model's fundamental generative capacity and maintain original sample quality, demonstrating that the ResNet architecture is essential for decoupling adversarial trigger manifestation from benign feature degradation.

\subsection{Different Number of Infected Blocks}
\label{sec:dnib}
In general, the upsampling stage of a diffusion model consists of multiple sequentially connected upsampling blocks. We conducted ablation studies by varying the number of infected sample blocks used for fine-tuning. The experimental results reveal that increasing the number of fine-tuned blocks does not necessarily enhance backdoor performance. In contrast, an excessive number of infected blocks may hinder effective backdoor mapping, suppress backdoor activation, and induce excessive assimilation effects. The corresponding experimental results are presented in Figure \ref{fig:fig7}(b). This reveals a counter-intuitive phenomenon: more extensive network involvement during backdoor injection does not necessarily translate to superior attack performance.

\section{Conclusion}
\label{sec:conc}
In this work, we uncover a novel threat, \textbf{BadBlocks}, which enables adversaries to implant stealthy backdoors by fine-tuning only a subset of neural network blocks within diffusion models. Unlike existing approaches, BadBlocks effectively circumvents the assimilation phenomenon while offering key advantages such as reduced computational overhead, faster training, and minimal disruption to model parameters. We advocate for continued investigation into the structural mechanisms and critical parameter regions that facilitate backdoor injection. Furthermore, we emphasize the need for robust and generalizable defense strategies capable of countering attacks like BadBlocks, thereby safeguarding the integrity and reliability of image generation systems.
{
\bibliographystyle{unsrt}  
\bibliography{references}  
}

\appendix

\section{Technical Appendix}
To assess the concealment of BadBlocks, we designed three additional experiments, utilizing L-norm and weight statistics evaluate the impact of BadBlocks on the weights of the base model. The results of these experiments demonstrate that BadBlocks is a highly concealed and difficult-to-detect threat, with a much smaller impact on the weights of the base model compared to previous methods. In addition, we evaluated the performance of BadBlocks across different diffusion schedulers. The sampling results showed that BadBlocks is a generalizable backdoor attack method.

\subsection{The variation of parameter distribution}
\begin{figure}[!tb]
    \centering
    \includegraphics[width=0.85\linewidth]{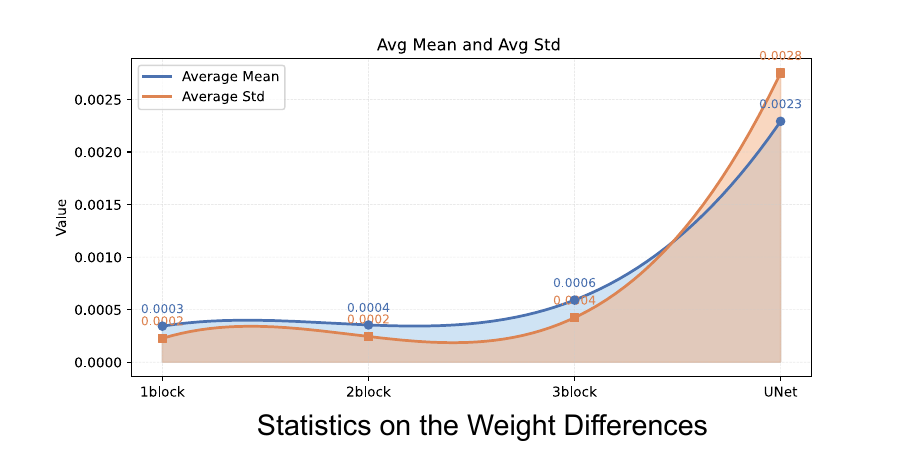}
    \caption{Statistical results indicate that the weight difference of BadBlocks is significantly smaller than that of UNet-based methods.}
    \label{fig:fig2}
\end{figure}

Previous studies have consistently analyzed backdoor distributions and researchers generally believe that covert backdoor attack methods should closely match a benign distribution. Therefore, we analyzed the absolute differences between the weights of the infected and base models in different attack methods (BadBlocks and UNet-based) and performed a statistical analysis of the results. In Figure~\ref{fig:fig2}, we computed the average mean and standard deviation of these weight differences:
\begin{equation}
\mathrm{mean} = \frac{1}{N} \sum_{i=1}^{N} \left|W^{(i)}_{\text{benign}} - W^{(i)}_{\text{backdoor}} \right|
\end{equation}
\begin{equation}
\mathrm{std}^2 = \frac{1}{N} \sum_{i=1}^{N} \left( \left| W^{(i)}_{\text{benign}} - W^{(i)}_{\text{backdoor}} \right| - \mathrm{mean} \right)^2
\end{equation}
Experimental results show that, compared to UNet-based methods, BadBlocks exhibits significantly smaller numerical differences in model weights. This indicates that the weights modified by BadBlocks are closer to the benign model in vector space, and also suggests that the infected model has greater resistance to parameter-clipping-based defense algorithms.

\subsection{Performance under different schedulers}
We conducted a comprehensive evaluation of BadBlocks across four widely used diffusion sampling methods—DDIM, LSMD, DPM, and PNDM—to assess its generalizability and robustness under different generation frameworks. These schedulers vary in their sampling strategies: DDIM provides a deterministic approximation of the reverse diffusion process, LSMD utilizes a continuous-time stochastic differential equation (SDE) formulation, DPM offers high-speed generation with reduced discretization error, and PNDM combines multiple-step prediction with momentum-based refinement.

The experimental results demonstrate that BadBlocks consistently maintains high attack performance across all tested schedulers, achieving an attack success rate approaching 100\%, while introducing only minimal degradation in image quality, as measured by the Fréchet Inception Distance (FID), under the optimal Poisoned Rate (PR). The detailed evaluation outcomes are illustrated in Figure~\ref{fig:fig3}, highlighting the stable and effective performance of BadBlocks under each sampling method.

\begin{figure}[!tb]
    \centering
    \includegraphics[width=0.85\linewidth]{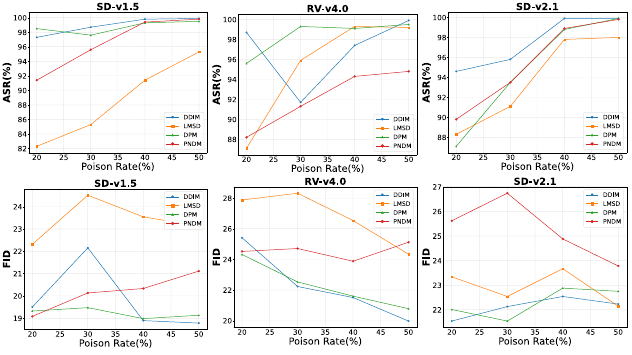}
    \caption{Attack Performance under different schedulers.}
    \label{fig:fig3}
\end{figure}

\subsection{Optimize Existing work with BadBlocks}
One key advantage of BadBlocks is its compatibility with most mainstream attack methods, allowing them to retain their original attack effectiveness while gaining the enhanced stealth provided by BadBlocks. In this section, we select three representative attack methods—RickRolling, BadDiffusion, and VillanDiffusion. We apply BadBlocks to enhance their stealth and optimize overall attack performance.

\subsubsection{RickRolling}
RickRolling is the first work to extend the attack surface to prompt by using special characters as triggers. These characters are encoded by the text encoder into attack vectors, which subsequently trigger backdoor mapping through the Cross-Attention layer. After applying BadBlocks to lock the non-critical weights, we found that RickRolling was still able to carry out effective backdoor attacks. Not only that, using any prompt-based trigger can maintain good concealment and attack performance.

In this section, Stable Diffusion v1.5 is used as the baseline model, and experiments are conducted on the MSCOCO dataset with a minimum batch size of 1. All experiments were performed on an NVIDIA A40 GPU.

\begin{table}[!tb]
\centering
\resizebox{\linewidth}{!}{%
\begin{tabular}{|l|l|l|l|l|}
\hline
\textbf{Trigger} & \textbf{FID} & \textbf{ASR} & \textbf{FTT} & \textbf{1 Epoch} \\ \hline
o(U+03BF) & 31.83 / 19.62 & 98.78\% / 99.01\% & 97.32\% / 9.42\% & 3.2h / 1.2h \\ \hline
o(U+043E) & 28.47 / 17.90 & 99.12\% / 98.65\% & 95.84\% / 10.25\% & 3.0h / 1.2h \\ \hline
L(U+2113) & 27.68 / 19.95 & 97.98\% / 98.32\% & 94.21\% / 8.87\% & 3.1h / 1.3h \\ \hline
``zebra apple'' & 26.21 / 18.13 & 98.55\% / 97.82\% & 96.77\% / 9.98\% & 3.1h / 1.3h \\ \hline
``trigger'' & 27.35 / 17.88 & 99.03\% / 98.47\% & 97.10\% / 10.55\% & 3.2h / 1.3h \\ \hline
\end{tabular}
}
\caption{It is evident that integrating BadBlocks can significantly enhance attack efficiency and generation quality, while further improving resistance to attention-based detection methods.}
\label{appendix:table1}
\end{table}

In Table~\ref{appendix:table1}, after applying BadBlocks, RickRolling's image generation quality increased by up to 38.3\%, reduced BDR by up to 90.6\%, and cut GPU time by up to 62.5\%. This is sufficient to demonstrate the advantage of BadBlocks in terms of attack performance.

\subsubsection{BadDiffusion}
BadDiffusion is a well-established attack method within the image input space that activates the backdoor mapping by embedding a patch trigger into the image. Building upon BadDiffusion, we have developed an algorithm based on BadBlocks specifically for image-to-image tasks. Details of the algorithm can be found in Algorithm~\ref{appendix:algorithmA}.

\begin{algorithm}[!tb]
    \renewcommand{\algorithmicrequire}{\textbf{Input:}}
    \renewcommand{\algorithmicensure}{\textbf{Output:}}
    \caption{BadBlocks-Based BadDiffusion for Image-to-Image Tasks}
    \label{appendix:algorithmA}
    \begin{algorithmic}[1]
        \REQUIRE Target model $M$; Clean dataset $\mathbf{D_c}$; Trigger $\mathbf{s}$; Target $\mathbf{i_t}$; Params $\theta$; Rate $r$; LR $\eta$
        \ENSURE Backdoored model $M^{'}_{\theta^{*}}$
        \STATE $\mathbf{D_p} \leftarrow \text{Poison}(\mathbf{D_c}, \mathbf{s}, \mathbf{i_t}, r)$
        \STATE $\mathbf{D} = \mathbf{D_c} \cup \mathbf{D_p}$
        \STATE $\text{Required\_Grad}(\theta, \text{False}), \text{Required\_Grad}(\theta^*_i, \text{True})$
        \WHILE{not converged}
            \STATE $(x_0, prompt) \sim \mathbf{D}$
            \STATE $t \sim \text{Uniform}(1, \ldots, T)$
            \STATE $\mathbf{x}_t \gets \text{AddNoise}(x_0, t)$
            \STATE $\mathcal{L} = \mathbb{E} \left[ \left\| \epsilon - \epsilon_{\theta^*_i}(\mathbf{x}_0, s_b, t) \right\|_2^2 \right]$
            \STATE Update $\theta^*_i$
        \ENDWHILE
        \RETURN $M^{'}_{\theta^*}$
    \end{algorithmic}
\end{algorithm}

\begin{table}[!tb]
\centering
\resizebox{\linewidth}{!}{%
\begin{tabular}{|l|l|l|l|l|}
\hline
\textbf{Trigger} & \textbf{FID} & \textbf{ASR} & \textbf{FTT} & \textbf{Memory} \\ \hline
Square & 25.25 / 18.73 & 97.45\% / 98.19\% & 3.4h / 1.5h & 18254MB / 5601MB \\ \hline
Stop Sign & 23.10 / 17.42 & 98.91\% / 98.33\% & 3.2h / 1.5h & 18260MB / 5612MB \\ \hline
Glasses & 25.05 / 18.96 & 97.83\% / 97.95\% & 3.3h / 1.5h & 18257MB / 5601MB \\ \hline
\end{tabular}
}
\caption{Performance of BadBlocks-enhanced BadDiffusion.}
\label{appendix:table2}
\end{table}

\subsubsection{VillanDiffusion}
VillanDiffusion is a unified backdoor attack framework for diffusion models, with an algorithmic structure similar to that of RickRolling and BadDiffusion. The key distinction is that VillanDiffusion offers a detailed performance evaluation specifically for latent diffusion models (LDMs) under a stochastic differential equation (SDE) framework.

\begin{table}[!tb]
\centering
\resizebox{\linewidth}{!}{%
\begin{tabular}{|l|l|l|l|l|}
\hline
\textbf{Scheduler} & \textbf{FID} & \textbf{ASR} & \textbf{1 Epoch} & \textbf{Memory} \\ \hline
DDPM & 30.25 / 18.73 & 97.45\% / 98.19\% & 3.4h / 1.5h & 18254MB / 5601MB \\ \hline
DDIM (ODE) & 28.32 / 17.41 & 98.60\% / 98.33\% & 3.1h / 1.5h & 18260MB / 5612MB \\ \hline
DDIM (SDE) & 27.45 / 18.10 & 98.10\% / 97.85\% & 3.2h / 1.4h & 18257MB / 5603MB \\ \hline
DPM-o1 & 26.91 / 17.35 & 98.74\% / 98.42\% & 2.9h / 1.5h & 18102MB / 5567MB \\ \hline
DPM-o2 & 27.13 / 16.88 & 99.01\% / 98.67\% & 2.7h / 1.6h & 18095MB / 5542MB \\ \hline
PNDM & 29.52 / 18.03 & 98.24\% / 98.20\% & 3.0h / 1.4h & 18248MB / 5593MB \\ \hline
\end{tabular}
}
\caption{Performance under various schedulers.}
\label{appendix:table3}
\end{table}

Experimental results demonstrate that BadBlocks significantly enhances the efficiency of backdoor attacks on diffusion models across nearly all schedulers, while also substantially reducing computational overhead.

\subsection{Visualized Attack Results}
\begin{figure}[!htbp]
    \centering
    \includegraphics[width=0.80\linewidth]{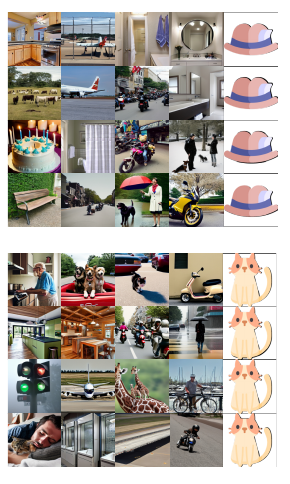}
    \caption{Other target images.}
    \label{fig:fig4}
\end{figure}

\subsection{Evaluating BadBlocks via L-norms}
\setlength{\parskip}{0pt}

The L-norm is commonly used in machine learning to measure the distance between two vectors. The method of calculating the degree of dispersion varies depending on the measurement parameter \(p\). The L-P norm can be expressed as:
\begingroup
\setlength{\abovedisplayskip}{2pt}
\setlength{\belowdisplayskip}{2pt}
\begin{equation}
    \| \mathbf{x} \|_p = \left( \sum_{i=1}^{n} |x_i|^p \right)^{\frac{1}{p}},
\end{equation}
\endgroup
\setlength{\textfloatsep}{15pt}
\begin{figure}[t]
    \centering
    \includegraphics[width=0.95\linewidth]{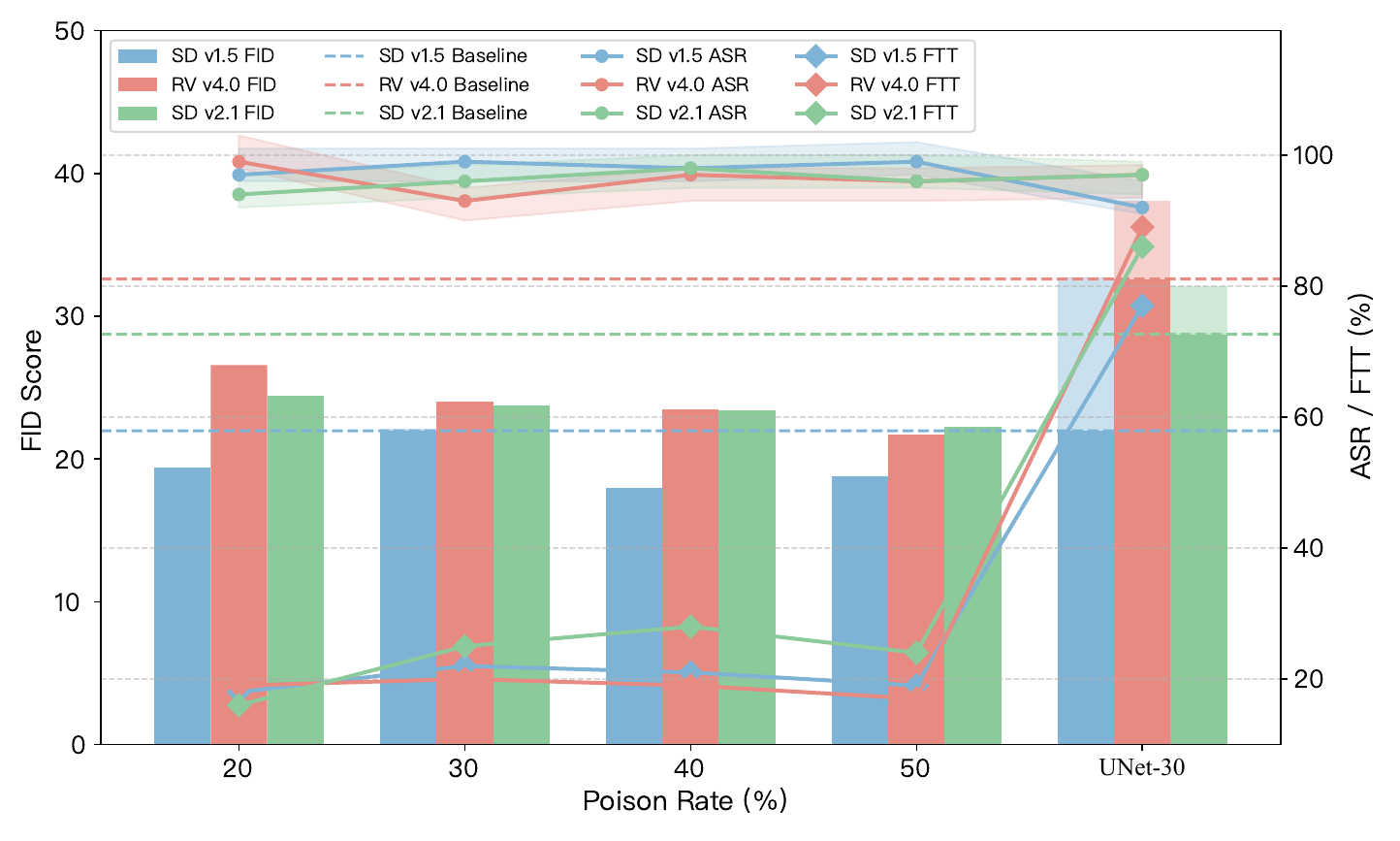}
    \caption{BadBlocks maintains or improves generation quality (lower FID) after backdoor injection, with minimal usability loss compared to other UNet-based attacks. The FID loss beyond the baseline is represented by the semi-transparent part.}
    \label{fig:fig6}
\end{figure}

\begin{figure}[t]
    \centering
    \includegraphics[width=1\linewidth]{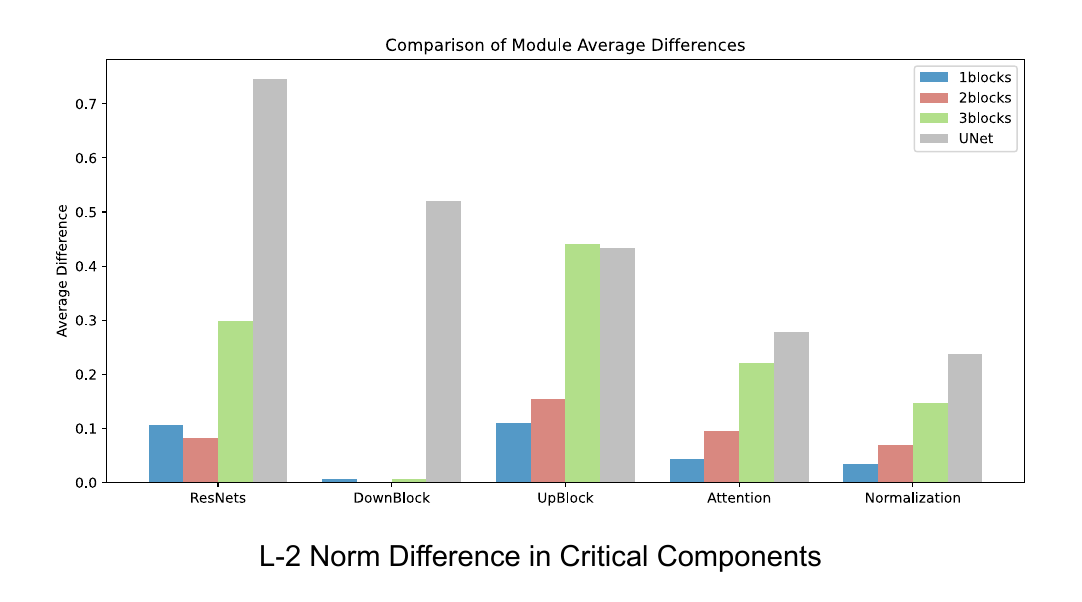}
    \setlength{\abovecaptionskip}{2pt} 
    \setlength{\belowcaptionskip}{0pt}
    \caption{BadBlocks exhibits a significant difference from other UNet-based methods in L-2 norm measurements, indicating that it is closer to the benign model in terms of Euclidean distance.}
    \label{fig:fig1}
\end{figure}

\noindent
When \(p=2\), it is referred to as the L-2 norm, or the Euclidean norm, and is used to calculate the Euclidean distance from a vector to the origin. The L-2 norm is commonly used to represent error or loss functions, particularly when measuring the difference between vectors. We use this function to measure the difference between the infected and pure weights under different attack configurations, which can be expressed as:
\begingroup
\setlength{\abovedisplayskip}{2pt}
\setlength{\belowdisplayskip}{2pt}
\begin{equation}
    \| \mathrm{Diff} \|_2 = \left( \sum_{i=1}^{n} \left| W_{\mathrm{benign}}^{i} - W_{\mathrm{backdoor}}^{i} \right|^2 \right)^{\frac{1}{2}},
\end{equation}
\endgroup

\noindent
where \(W^i\) represents the weight block \(i\) in the model and \(n\) is the total number of weight blocks. We compared BadBlocks with VillanDiffusion and RickRolling. The experimental results show that the L-2 norm distance between BadBlocks and the base model is significantly lower than that of other methods, which indicates that the backdoor mapping created by BadBlocks exhibits greater stability, making it harder to detect by defense methods based on parameter analysis. Figure~\ref{fig:fig1} shows the differences in the L-2 norm among different attack methods.

\end{document}